\documentclass {emulateapj}
\usepackage{apjfonts}
                                                                                
\newcommand{\mum}{\ifmmode{\rm \mu m}\else{$\mu$m}\fi}
 
\slugcomment{Accepted for publication in the Astrophysical Journal Supplement Series\hfill}
                                                                                
\begin{document}

\title{Spitzer Infrared Spectrograph\footnotemark[1] \ 
   Observations of M, L, and T Dwarfs}

\footnotetext[1]{The
IRS was a collaborative venture between Cornell University and Ball 
Aerospace Corporation funded by NASA through the Jet Propulsion 
Laboratory and the Ames Research Center}

\author{T.~L.~Roellig,\altaffilmark{2} J.~E.~Van~Cleve,\altaffilmark{3} 
G.~C.~Sloan,\altaffilmark{4} J.~C.~Wilson,\altaffilmark{5} D.~Saumon,\altaffilmark{6} 
S.~K.~Leggett,\altaffilmark{7} 
M.~S.~Marley,\altaffilmark{8} M.~C.~Cushing,\altaffilmark{9, 10} J.~D.~Kirkpatrick,\altaffilmark{11} 
A.~K.~Mainzer,\altaffilmark{12} and J.~R.~Houck\altaffilmark{4}}

\altaffiltext{2}{NASA Ames Research Center, MS 245-6, Moffett
  Field, CA 94035-1000, Thomas.L.Roellig@nasa.gov}
 \altaffiltext{3}{Ball Aerospace and Technologies Corp., 1600 Commerce St., Boulder, CO 80301}
  \altaffiltext{4}{Cornell University, Astronomy Department, Ithaca, NY 14853-6801}
 \altaffiltext{5}{Astronomy Bldg.,
University of Virginia,
530 McCormick Rd.,
Charlottesville, VA 22903}
 \altaffiltext{6}{ Los Alamos National Laboratory,
Applied Physics Division,
Mail Stop F699,
Los Alamos, NM 87545}
\altaffiltext{7}{ Joint Astronomy Centre, University Park, Hilo, HI 96720}
\altaffiltext{8}{NASA Ames Research Center, MS 245-3, Moffett
  Field, CA 94035-1000}
\altaffiltext{9}{ SETI Institute, NASA Ames Research Center, MS 245-3, Moffett
  Field, CA 94035-1000}
\altaffiltext{10 }{Spitzer Fellow}
\altaffiltext{11}{ Infrared Processing and Analysis Center, MS 100-22, California Institute of Technology, Pasadena, CA 91125}
\altaffiltext{12 }{Jet Propulsion Laboratory, Mailstop 169-506, 4800 Oak Grove Dr., Pasadena, CA 91109}

\begin{abstract}
We present the first mid-infrared spectra of brown dwarfs, together 
with observations of a low-mass star.  Our targets are the M3.5 dwarf 
GJ 1001A, the L8 dwarf DENIS-P J0255$-$4700, and the T1/T6 binary system 
$\epsilon$ Indi Ba/Bb.  As expected, the mid-infrared spectral 
morphology of these objects changes rapidly with spectral class due to 
the changes in atmospheric chemistry resulting from their differing 
effective temperatures and atmospheric structures.  By taking advantage 
of the unprecedented sensitivity of the Infrared Spectrograph 
on the Spitzer Space Telescope we have detected the 
7.8~\micron\ methane and 10~\micron\ ammonia bands for the first time in 
brown dwarf spectra.
\end{abstract}


\section{Introduction} 

Within the past few years the number of known brown dwarfs  
has increased dramatically, with both the Sloan Digital Sky 
Survey \citep[SDSS;][]{yor00} and Two Micron All-Sky Survey  
\citep[2MASS;][]{skr97} accounting for the 
bulk of the discoveries. These objects have been identified based on 
broad-band photometry with follow-up optical and near-infrared 
spectroscopy.  The integrated data from these 
investigations has yielded important information about the
variation of the chemical composition and of the effects of condensates
in brown dwarfs in term of their spectral types \citep[e. g.][]{kir00, 
leg01, bur02, mcl03, kna04}.

Spectroscopic observations in the mid-infrared yield important 
additional information \citep{sau03}, but due to 
the very low infrared brightness of these objects and the effects of 
the Earth's atmospheric absorption these measurements have been 
difficult or impossible from ground-based observatories. With the 
launch of the Spitzer Space Telescope \citep{wer04} and 
the onboard Infrared Spectrograph \citep[IRS;][]{hou04} 
on 2003 August 25, we now have an observatory with sufficient 
sensitivity to undertake these studies.  Accordingly, the Science Team 
of the IRS has organized a comprehensive observational program of M, L, 
and T dwarfs, covering the range from the earliest M dwarfs to the 
latest T dwarfs.  We present here the first results from this program: 
observations of the M3.5 star GJ 1001A (LHS 102A), the L8 dwarf DENIS-P J0255$-$4700 
(hereafter J0255$-$4700), and the T1/T6 binary system $\epsilon$ Indi 
Ba/Bb.

\section{Observations and Flight Data Reduction} 

The three objects described here were observed with the IRS 
instrument on the Spitzer Space Telescope as part of one of the 
major IRS Science Team Guaranteed Time Observing programs.  The 
IRS is capable of low-and moderate-resolution spectroscopy, with 
only data from the low-resolution modules being reported here. 
Each of the IRS low-resolution modules is fed by two long slits, 
one used for first order and the second used for second order. The IRS low-resolution 
modules have slit lengths of 54\farcs 6 and 151\farcs 3 for both orders in the 
short-low (SL) and long-low (LL) modules, respectively.  
The spectral resolution of the SL module as determined 
by the slit width is $\Delta\lambda$ = 0.06 \micron\ for second order 
and  $\Delta\lambda$ = 0.12 \micron\ for first order.  The spectral 
resolution of the LL module as determined by the slit width 
is $\Delta\lambda$ = 0.17 \micron\ for second order and 
$\Delta\lambda$ = 0.32 \micron\ for first order.  The slit widths of both of the IRS
low-resolution modules are Nyquist sampled by the 
 pixel pitch of their infrared arrays.
\citet{hou04} describe the design, performance,
  and operation of the IRS in more detail.

Table 1 provides a summary of the observations reported here.
As all of these objects have relatively large proper motions, the IRS 
blue peakup array was used to locate the current positions of these 
objects and to reposition them into the IRS slits with an estimated 
radial uncertainty of approximately 0\farcs4, (1$\sigma$).  Each object was 
observed at two locations in 
each of the IRS low-resolution slits, with the positions located at approximately
the 1/3 and 2/3 positions along the long axis of the slits.

The standard IRS data reduction pipeline version
S9.1 at the Spitzer Science Center processed the data. 
Stellar standards were also observed to calibrate 
the spectra and remove any spectral artifacts which may have survived the
processing; the results 
reported here were corrected with the A dwarf standard star $\delta$ UMi.
The zodiacal and galactic backgrounds were subtracted before the spectral extractions
using observations when the target was located in the alternate slit. Finally, the 
spectra were extracted from the sky-subtracted two-dimensional array images 
using an early version of the SMART IRS data-reduction software 
package being developed at Cornell University \citep{hig04}.

Figure 1 shows the resulting IRS low-resolution spectra for the three 
objects reported here.  At this early stage in the Spitzer mission, our
estimated uncertainty in the absolute photometry is $\pm25\%$, but the
uncertainty in the spectral shape is much less than this.  The caption for 
Figure 1 gives the
uncertainties in the spectral shapes which were estimated from a combination 
of the measured noise in the individual reads of the pixels in the 
detector arrays and the variation between the spectra extracted from 
the observations at the two slit positions. These uncertainties therefore 
incorporate some measure 
of the residual systematic uncertainties and spectral flat-fielding 
errors.  Uncertainties in the absolute flux calibration were estimated 
in different ways for the three objects and are described below.  

\begin{deluxetable}{lll} 
\tablecolumns{3}
\tablewidth{0pt}
\tablenum{1}
\tablecaption{Table of Observations \label{Tbl1}}

\tablehead{ \colhead{Object} & \colhead{Observation Dates} &
  \colhead{Spitzer AOR Key}  \\ \colhead{} & \colhead{Integration Time} &
  \colhead{} }

\startdata
GJ 1001A & 2003 December 16 & 4190464 \\
\\
         & SL2   480 seconds \\
         & SL1   480 seconds \\
         & LL2   480 seconds \\
         & LL1   960 seconds \\
\\
DENIS-P J0255$-$4700 & 2004 January 7 & 4192000 \\
\\
                   & SL2  480 seconds \\
                   & SL1  480 seconds \\
                   & LL2  480 seconds \\
                   & LL1  960 seconds \\
\\
$\epsilon$ Indi Ba/Bb & 2003 September 26 & 6625792\tablenotemark{a} \\
\\
                      & SL2  56 seconds\\
                      & SL1  56 seconds \\
\enddata
\tablenotetext{a}{Observations of the binary system $\epsilon$ Indi 
Ba/Bb \citep{sch03, mcc04} were taken during 
the Spitzer In-Orbit Checkout and Science Validation period which 
accounts for it being observed only by the IRS Short-Low module and 
for such a short integration time.  Additional 
observations of this object will take place later in the Spitzer mission.}
\end{deluxetable}

\begin{figure} 
\includegraphics[height=4.5in]{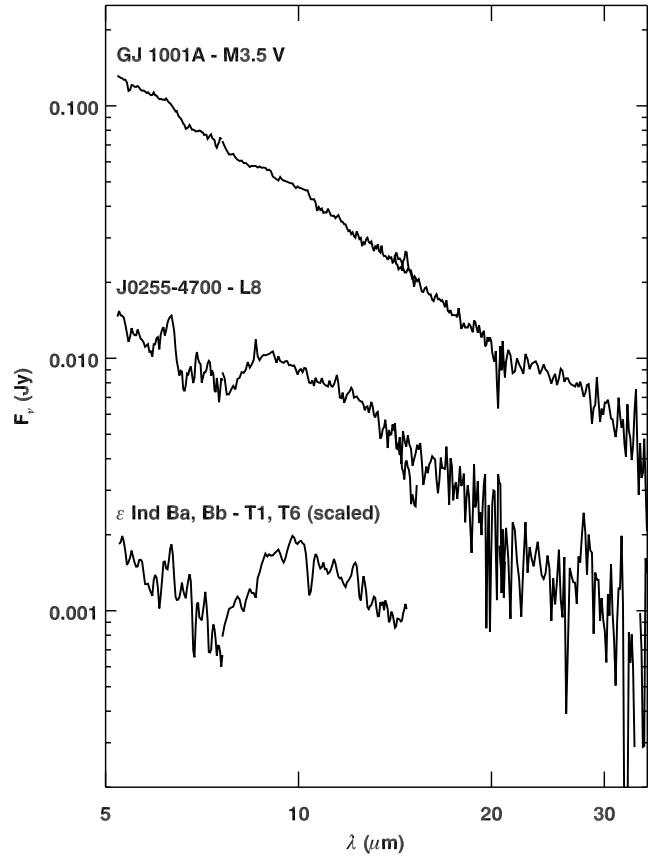}
\caption{The observed IRS spectra of GJ 1001A, J0255$-$4700, and the 
$\epsilon$ Indi Ba/Bb system.  The fluxes for $\epsilon$ Indi Ba/Bb 
have been divided by a factor of 10 for clarity.  The average 
1$\sigma$ errors in the data points are as follows: for GJ 1001A, 2.60 mJy
from 5.3 to 14.1 \micron\ and 1.61 mJy from 14.1 to 35.0 \micron; for 
J0255$-$4700 they are 0.73 mJy from 5.3 to 14.1 \micron\ and 1.00 mJy from 
14.1 to 35.0 \micron; for the $\epsilon$ Indi Ba/Bb system they are 2.26
mJy from 5.3 to 15.3 \micron.}
\end{figure}

\begin{figure} 
\includegraphics[height=4.5in]{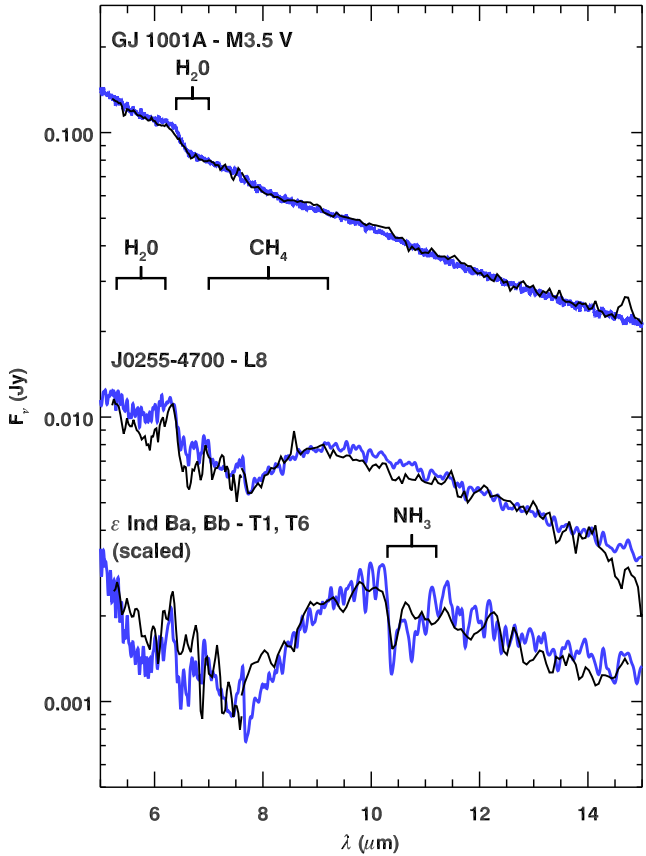}
\caption{The observed IRS spectra of GJ 1001A, J0255$-$4700, and the $\epsilon$ 
Indi Ba/Bb system compared with model spectra.  The fluxes for $\epsilon$ Indi 
Ba/Bb have been divided by a factor of 10 for clarity. The observed data are denoted 
by the solid black lines while the model spectra are in the thicker blue lines.  The 
various model parameters for each of the spectra are given in the text.  The most 
noticeable H$_2$O, CH$_4$, and NH$_3$ bands are marked, although weaker 
H$_2$O absorption extends from the blue edge of the figure to 7.0 \micron, and 
NH$_3$ absorption in $\epsilon$ Indi Ba/Bb extends from 8.5 \micron \  to the 
red edge of the figure. In the case of GJ 1001A the entire displayed spectrum is 
dominated by water absorption, while in J0255$-$4700 water is dominant longward of 
9 \micron. In a T1 (the brightest component of $\epsilon$ Indi B), NH$_3$ is present 
but not strong enough to be visible in the 12--15 \micron\ region, so that longward of 
12 \micron \ the spectral morphology is still primarily due to water absorption. The 
strong 7.8 \micron\ absorption band from CH$_4$ is evident in both of the cooler objects.}
\end{figure}

\section{Discussion} 

As the atmospheres of brown dwarfs cool with time, their spectral signatures
reflect a progression of changes in their atmospheric chemical equilibrium
and condensate structure.  The three dwarfs discussed here provide snapshots
of this progression.   In an M dwarf such as GJ 1001A, the elements O, C, and N are
predominantly found in $\rm H_2$O, CO, and $\rm N_2$ and the atmosphere is too
warm for condensation of solids (\cite{all95, lod99}).  As the effective
temperature ($T_{eff}$) falls, a variety of condensates form in the atmosphere, most notably
iron and silicates.  These condensates are not expected to be well-mixed
through the atmosphere, but rather be found in relatively thin, discrete
cloud layers overlying the condensation level (\cite{ack01, mar02, tsu02, woi04}). 
 As the  $T_{eff}$ falls to that of a late L dwarf, such as
J0255$-$4700, the cloud layer is optically thick and affects either directly (as
a major opacity source) or indirectly (by altering the atmospheric
temperature/pressure profile) all spectral regions.  The exact spectral
signature of the cloud depends both on its vertical thickness and the particle
size distribution of the condensates.  In addition as the atmosphere cools
chemical equilibrium begins to favor first $\rm CH_4$ over CO and then
$\rm NH_3$ over $\rm N_2$ (\cite{tsu64, feg96, bur99, lod99, lod02, bur01}).  The strong mid-infrared 
bands of these
molecules begin to challenge the domination of water opacity, at least in
some spectral regions, by the late L's.  By the early to mid T dwarfs, such
as the $\epsilon$ Indi pair, the condensate cloud is forming quite deep in
the atmosphere.  In the relatively clear, cool atmosphere above the cloud,
chemical equilibrium begins to strongly favor $\rm CH_4$ and $\rm NH_3$ and
their spectral features, along with particularly strong bands of water, grow
in prominence (\cite{mar96, bur97, all01, bur03}).

Figure 1 shows the resulting pronounced changes in the mid-infrared spectral morphology 
between the three objects.  The spectra and comparisons with synthetic 
spectra are considered for each of the objects in turn.

\subsection{GJ 1001A}

GJ 1001A is classified as an M3.5 V star \citep{haw96} and is part of a binary system 
with the L5 dwarf GJ 1001B
\citep{gol99}.  The distance between the primary and secondary 
is large enough (18\arcsec) that there is no contamination of the 
primary's spectrum by stray light from the secondary.  Using optical
and near-infrared spectra, Leggett et al. (2002) find a best fit to 
this object with $T_{eff}$ = 3200 K and ${\rm log}  \ g = 5$ using a mixing
length parameter of 2 in the AMES-dusty models of Allard et al. (2001).
Figure 2 compares this same model, smoothed to a spectral resolution 
of $R\sim 400$, with the IRS data.  The 6.5 \micron\ feature seen in
the IRS spectrum also appears in the model and arises from H$_{2}$O
opacity.

For the object GJ 1001A the near-infrared observations \citep{leg02} can 
be used with the model and the resulting predicted photometric flux 
compared with our observations.  Although the shape of the model 
spectrum is an excellent match to the IRS data, the model fluxes derived 
from the near-infrared observations had to be scaled by a factor of 
0.87 to fit the observed data.  The near-infrared flux 
levels are consistent with the $T_{eff}$ and radius derived from the luminosity 
if the dwarf is older than 0.15 Gyr \citep{leg02, bar98}.  

\subsection{J0255$-$4700}

J0255$-$4700 is classified optically as an L8 
dwarf.\footnote{http://spider.ipac.caltech.edu/staff/davy/ARCHIVE}
Unfortunately the trigonometric parallax distance of this object has 
not been measured. Considerable structure can be seen in the IRS 
spectrum and to interpret these results we have compared the IRS data 
with the model described by \citet{mar02} and Saumon et al. (2003), with 
T$_{eff}$ = 1400 K, ${\rm log}  \ g = 4.5$  in 
Figure 2.  In this model atmospheric condensate clouds are treated as described 
by \citet{ack01} with a sedimentation efficiency 
parameter $f_{sed} = 3$.  The model has been
normalized to match the observed flux in the 8.6 -- 9.0 \micron \ 
region. Models that include condensate sedimentation 
in the chemical equilibrium  but ignore cloud opacity require 
unrealistically high T$_{eff}$ to acceptably fit the data and these 
are not shown here.  We judge the excellent agreement of the cloudy 
model to the J0255$-$4700 data as further evidence for finite-thickness 
silicate and iron clouds in the observable atmospheres of late L dwarfs.
As can be seen in the figure the correlation between the model 
spectrum and the IRS data is excellent, although there is a slight 
discrepancy between the data and model predictions in the H$_2$O 
feature at 5.5 -- 6.5 \micron.  In addition there is a broad shallow dip 
in the IRS data compared with the model between 9 and 11 \micron \ that may be
due to a silicate cloud feature. 
The IRS spectrum shows the first detection of the 7.8 \micron\ CH$_4$ band 
in a brown dwarf.  $\rm CH_4$ was
detected at 3.3 \micron\  in L dwarfs by \citet{nol00}
and since the $7.8\,\rm \mu m$ band is of comparable strength to the $3.3\,\rm
\mu m$ fundamental, it is not surprising to find this mid-infrared signature of
atmospheric methane in an L dwarf. Models with (\cite{sau03}) and without (\cite{bur97, all01}) dust opacity also anticipated this feature in the L dwarf  T$_{eff}$ range along with the water features noted above.  Methane absorption is even more obvious in the 
spectrum of the $\epsilon$ Indi B system discussed below.

\citet{cre04} used OSCIR at Gemini South to
image J0255$-$4700 in the $N$-band ($\sim 8.1$ -- 13.4 \micron)
and three narrowband filters centered at 8.8, 10.3 and
11.7 \micron.  Flux density detections were obtained
in the $N$ and 8.8 \micron\ bands, with upper limits for the
10.3 and 11.7 \micron\ bands.  Despite the estimated 25\%
uncertainty in our absolute flux calibration, the IRS spectrum
cannot be scaled to match both their $N$-band and their
8.8 \micron\ flux density measurements.   \citet{cre04}
note that the $N$-band flux density is a factor of 2 to 3
greater than the narrowband values and discuss possible causes of
this discrepancy. Our data can rule out two of their suggestions - the presence 
of peculiar emission or absorption lines.
Figures 1 and 2 show that there are no strong emission features
in the $N$ bandpass and also that the narrow bandpasses do
not fall in regions of strong absorption.  While it is
difficult to reconcile all of the Creech-Eakman et al.
photometry with our IRS spectrum, our data can be brought
into approximate agreement with the three narrowband
measurements if the spectrum is scaled to the lower limit
of the 8.8 $\mu$m OSCIR detection.  This may indicate that
the $N$-band measurement is in error.

\subsection{$\epsilon$ Indi Ba and Bb}

The $\epsilon$ Indi Ba and Bb system consists of a T1 and T6 dwarf 
separated by 0\farcs 732 \citep{mcc04}.  The IRS does 
not have the spatial resolution to separate the two objects so the 
measured spectrum is a composite of the two individual spectra.  
In this object we find the first evidence for 
NH$_3$ absorption in the spectrum between 10 and 11 \micron, although 
its presence in very cool brown dwarf atmospheres has been expected 
for some time \citep{tsu64, mar96, feg96, sau00, all01, bur03, sau03}.  
\citet{sau00} reported a weak detection of NH$_3$ in the near 
infrared spectrum of the T6 dwarf Gl 229B but this is the first
strong detection of NH$_3$ in a brown dwarf.  With the lower 
T$_{eff}$ for the components of this binary system the condensate clouds are 
predicted to be below the photosphere and thus have a less pronounced 
effect on the composite spectrum than they do on the L8 dwarf's spectrum.

A composite model spectrum provides an excellent match to the IRS 
data and was obtained as follows.  As with the model for J0255$-$4700 
we used the model described by \citet{mar02} and Saumon et al. (2003).  
From the estimates of $L_{bol}$ for each 
component \citep{mcc04} and the estimated age of $\epsilon$ 
Indi A \citep[0.8--2 Gyr;][]{lac99}, we use our brown dwarf 
cooling calculations \citep{mar04} to obtain ($T_{eff}$, log\ $g$, $R/R_{\sun}$) = 
(1250 K, 5.13, 0.094), (840 K, 4.89, 0.100) for the T1 and the T6 
dwarfs, respectively.  Selecting the closest cloudless synthetic 
spectra in our grid of models (1200 K, 5.0) and (800 K, 5.0) which 
are within the uncertainties of those parameters, and using the above
radii and the trigonometric parallax of $\epsilon$ Indi A  \citep{per97}, the 
absolute flux at the Earth of the T dwarf binary was obtained.  Note 
that in Figure 2, the composite model spectrum has not been 
normalized to the IRS data, so that in this case the predicted flux 
agrees with the IRS data to within the data errors.

\section{Summary and Conclusions} 

(1) We have taken advantage of the unprecedented sensitivity of the 
Infrared Spectrograph instrument on the Spitzer Space telescope to 
observe two brown dwarfs.  These are the first spectra in this 
wavelength range ever reported for this type of object.  The spectrum 
of an M dwarf is also reported here, providing a contrast between 
objects with masses above and below the Hydrogen-Burning Minimum Mass.

(2) The observed mid-infrared spectral morphologies of these three objects vary 
strongly with their spectral classes, reflecting the changes in gas chemistry driven 
by the temperature of their atmospheres.

(3) Model comparisons with the observed spectra of all three objects 
show good agreement, with only a few minor deviations.  Models 
with cloud opacities do a better job of reproducing the data from J0255$-$4700 
than do models without clouds.

(4) We report here the first positive detection of the 7.8 \micron methane band.  
We also report the first unambiguous detection of NH$_3$ in a brown dwarf atmosphere.

\section{Acknowledgments} 

We are pleased to thank the entire team of dedicated 
scientists, engineers, and managers that contributed to the 
development of the Spitzer observatory and the IRS instrument.  In 
particular, we would like to especially thank Larry Simmons and David 
Gallagher of JPL, and John Troeltzsch, Marty Huisjen, and John Marriott of Ball Aerospace for their leadership in this endeavor. 
This work is based on observations made with the Spitzer Space 
Telescope, which is operated by the Jet Propulsion Laboratory, 
California Institute of Technology under NASA contract 1407.  
Support for this work was provided by NASA's Office of Space Science. 
T. Roellig and M. Marley would like to acknowledge the support of the NASA Office of 
Space Sciences. D. Saumon's  work at LANL is supported by the United 
States Department of Energy under contract W-7405-ENG-36. M. Cushing is supported
by a Spitzer Fellowship.

\end{document}